\documentclass[aps,reprint,superscriptaddress]{revtex4-1}
\usepackage[usenames,dvipsnames]{xcolor}
\usepackage{graphicx}
\usepackage{amsmath}
\usepackage{amssymb}
\usepackage{esint}
\usepackage{amsfonts}
\usepackage{natbib}
\usepackage{psfrag}
\usepackage{wasysym}
\usepackage{float}
\usepackage{epstopdf}
\usepackage{color}
\usepackage[normalem]{ulem}
\usepackage{textcomp}
\usepackage{bm}

\renewcommand{\d}[2]{\frac{\id #1}{\id #2}} 
\newcommand{\overbar}[1]{\mkern 1.5mu\overline{\mkern-1.5mu#1\mkern-1.5mu}\mkern 1.5mu}
\newcommand{\id}{\mathrm{d}} 
\newcommand{\ii}{\mathrm{i}}
\newcommand{\e}{\mathrm{e}}

\newcommand{\pdd}[2]{\frac{\partial^2 #1}{\partial #2^2}} 

\begin{document}

\title{Dynamical Theory of the Inverted Cheerios Effect}

\author{Anupam Pandey}
\email[]{a.pandey@utwente.nl}
\affiliation{Physics of Fluids Group, Faculty of Science and Technology, University of Twente, P.O. Box 217, 7500AE Enschede, The Netherlands}
\author{Stefan Karpitschka}
\affiliation{Physics of Fluids Group, Faculty of Science and Technology, University of Twente, P.O. Box 217, 7500AE Enschede, The Netherlands}
\author{Luuk A. Lubbers}
\affiliation{Huygens-Kamerlingh Onnes Lab, Universiteit Leiden, P.O. Box 9504, 2300RA Leiden, The Netherlands}
\author{Joost H. Weijs}
\affiliation{Universit{\'e} Lyon, Ens de Lyon, Universit{\'e} Claude Bernard, CNRS, Laboratoire de Physique, F-69342 Lyon, France}
\author{Lorenzo Botto}
\affiliation{School of Engineering and Materials Science, Queen Mary University of London,
London E1 4NS, UK}
\author{Siddhartha Das}
\affiliation{Department of Mechanical Engineering, University of Maryland, College Park, MD 20742, USA}
\author{Bruno Andreotti}
\affiliation{Laboratoire de Physique et M{\'e}canique des Milieux Het{\'e}rog{\`e}nes, UMR 7636 ESPCI- CNRS, Universit{\'e} Paris- Diderot, 10 rue Vauquelin, 75005 Paris, France}
\author{Jacco H. Snoeijer}
\affiliation{Physics of Fluids Group, Faculty of Science and Technology, University of Twente, P.O. Box 217, 7500AE Enschede, The Netherlands}
\affiliation{Department of Applied Physics, Eindhoven University of Technology, P.O. Box 513, 5600MB
Eindhoven, The Netherlands}

\date{\today}

\begin{abstract}
Recent experiments have shown that liquid drops on highly deformable substrates exhibit mutual interactions. This is similar to the Cheerios effect, the capillary interaction of solid particles at a liquid interface, but now the roles of solid and liquid are reversed. Here we present a dynamical theory for this inverted Cheerios effect, taking into account elasticity, capillarity and  the viscoelastic rheology of the substrate. We compute the velocity at which droplets attract, or repel, as a function of their separation. The theory is compared to a simplified model in which the viscoelastic dissipation is treated as a localized force at the contact line. It is found that the two models differ only at small separation between the droplets, and both of them accurately describe experimental observations.
\end{abstract}

\pacs{}

\maketitle
\section{Introduction}

The clustering of floating objects at the liquid interface is popularly known as the Cheerios effect \cite{Vella2005}. In the simplest scenario, the weight of a floating particle deforms the liquid interface and the liquid-vapor surface tension prevents it from sinking \cite{Vella2015}. A neighboring particle can reduce its gravitational energy by sliding down the interface deformed by the first particle, leading to an attractive interaction. The surface properties of particles can be tuned to change the nature of interaction, but two identical spherical particles always attract \cite{Kralchevsky2000}. Anisotropy in shape of the particles or curvature of the liquid interface adds further richness to this everyday phenomenon \cite{loudet2005,botto2012,Ershov13}. Self-assembly of elongated mosquito eggs on the water surface provides an example of this capillary interaction in nature \cite{loudet2011}, while scientists have exploited the effect to control self-assembly and patterning at the microscale \cite{bowden1997,Lewan10,Furst11,Cavallaro13,Stamou2000}. 

\begin{figure*}
\includegraphics[width=175mm]{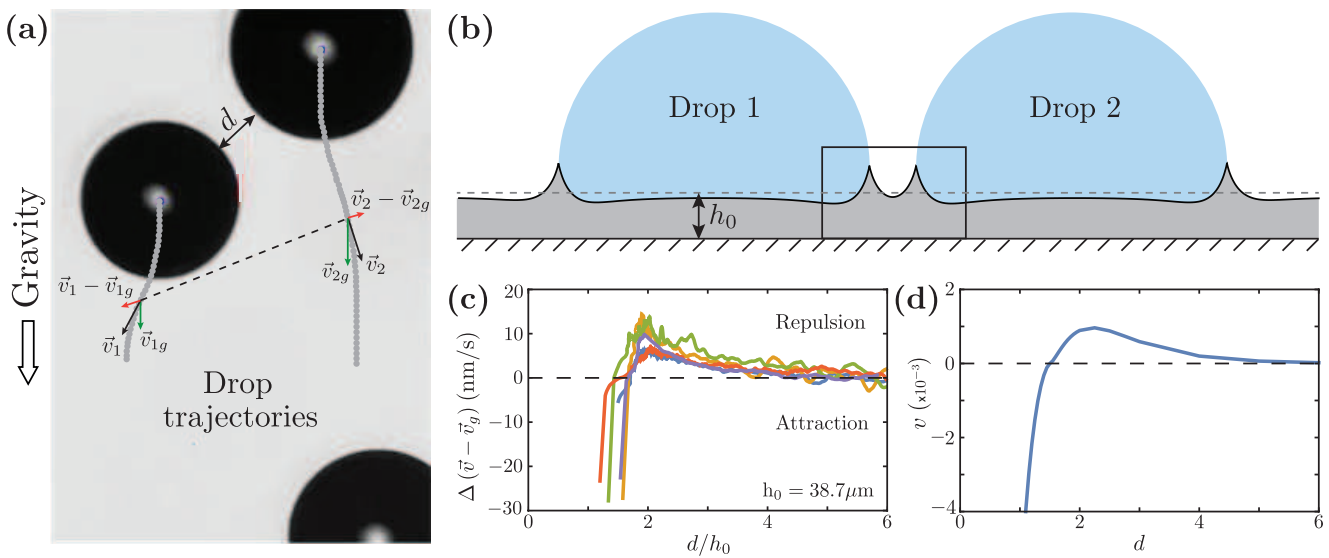}
\caption{Dynamics of the inverted cheerios effect. (a) Ethylene glycol drops of radius $R\simeq 0.5-0.8$ mm move down a vertically placed cross-linked PDMS substrate of thickness $h_0\simeq .04$ mm and shear modulus $G=280$ Pa under gravity. The arrows represent instantaneous drop velocity. Drop trajectories deviate from straight vertical lines due to a repulsive interaction between them. (b) A sketch of the cross-section of two drops on a thin, soft substrate. The region around two neighboring contact lines is magnified in fig.~\ref{int_fig}(a). (c) Relative interaction velocity as a function of the gap $d$ between drops for five pairs of drops. We use the convention that a positive velocity signifies repulsion and negative velocity signifies attraction. (d) A theoretical velocity-distance plot for power-law rheology.}
\label{expfig}
\end{figure*}

The concept of deformation-mediated interactions can be extended from liquid interfaces to highly deformable solid surfaces. Similar to the Cheerios effect, the weight of solid particles on a soft gel create a depression of the substrate, leading to an attractive interaction \cite{Maha2015,manoj14,manoj13}. Recently, it was shown that the roles of solid and liquid can even be completely reversed:  liquid drops on soft gels were found to exhibit a long-ranged interaction, a phenomenon called the \emph{inverted} Cheerios effect \cite{Karpitschka2016}. An example of such interacting drops is shown in fig.~\ref{expfig}(a). In this case, the droplets slide downwards along a thin, deformable substrate (much thinner as compared to drop size) under the influence of gravity, but their trajectories are clearly deflected due to a repulsive interaction between the drops. Here capillary traction of the liquid drops instead of their weight deforms the underlying substrate (cf fig.~\ref{expfig}(b)). The scale of the deformation is given by the ratio of liquid surface tension to solid shear modulus ($\gamma/G$), usually called the elasto-capillary length. Surprisingly, drops on a thick polydimethylsiloxane (PDMS) substrate (much thicker than the drop sizes) were found to always attract and coalesce, whereas for a thin substrate (much thinner than the drop sizes) the drop-drop interaction was found to be repulsive \cite{Karpitschka2016}. This interaction has been interpreted as resulting from the local slope of the deformation created at a distance by one drop, which can indeed be tuned upon varying the substrate thickness.

Here we wish to focus on the dynamical aspects of the substrate-mediated interaction. Both the Cheerios effect and the inverted Cheerios effect are commonly quantified by an effective potential, or equivalently by a relation between the interaction force and the particle separation distance \cite{Vella2005,Karpitschka2016}. However, the most direct manifestation of the interaction is the motion of the particles, moving towards or away from each other. In the example of fig.~\ref{expfig}(a), the dark gray lines represent drop trajectories and the arrows represent their instantaneous velocities.  When drops are far away, the motion is purely vertical due to gravity with a steady velocity $\vec{v}_g$. As the two drops start to interact, a velocity component along the line joining the drop centers develops ($\vec{v}_{i}-\vec{v}_{ig}$, $i=1,2$) that move the drops either away or towards each other. In fig.~\ref{expfig}(c), we quantify the interaction by $\Delta(\vec{v}-\vec{v}_g)=(\vec{v}_1-\vec{v}_{1g})-(\vec{v}_2-\vec{v}_{2g})$, representing the relative interaction velocity as a function of inter drop distance ($d$) for several interaction events. Even though the dominant behavior is repulsive, we find an attractive regime at small distances. Similar to the drag force on a moving particle at a liquid interface  \cite{drr2016}, viscoelastic properties of the solid resist the motion of liquid drops on it \cite{CGS96}. Experiments performed in the over damped regime allow one to extract the interaction force from the trajectories, after proper calibration of the relation for the drag force as a function of velocity \cite{Karpitschka2016}.

In this paper, we present a dynamical theory of elastocapillary interaction of liquid drops on a soft substrate. For large drops, the problem is effectively two dimensional (cf. fig.~\ref{expfig}(b)) and boils down to an interaction between two adjacent contact lines. By solving the dynamical deformation of the substrate, we directly compute the velocity-distance curve based on substrate rheology, quantifying at the same time the interaction mechanism and the induced dynamics. In sec.~\ref{sec:formulation} we set up the formulation of the problem and propose a framework to introduce the viscoelastic properties of the solid. In sec.~\ref{sec:results} we present our main findings in the form of velocity-distance plots for the case of very large drops. These results are compared to the previous theory for force-distance curves, and we investigate the detailed structure of the dynamical three phase contact line. Subsequently, our results are generalised in sec.~\ref{sec:radius} to the case of finite-sized droplets, and a quantitative comparison with experiments is given. 

\section{Dynamical Elastocapillary interaction}\label{sec:formulation}

\subsection{Qualitative description of the interaction mechanism}

Before developing a detailed dynamical theory of the inverted cheerios effect, we start by a qualitative discussion of the interaction mechanism. The problem of drop-drop interaction on soft substrates can be characterised by four length scales: the radius of the drop ($R$), the thickness of the substrate ($h_0$), the elastocapillary length ($\gamma/G$), and of course the gap separating the drops ($d$). In the case of drop size much larger than all other length scales: $R\gg \textup{max}(h_0,d,\gamma/G)$, one can neglect the curvature of the contact line and the wetting ridge becomes quasi two-dimensional as in fig.~\ref{expfig}(b). Furthermore the drop-drop interaction is limited to an interaction between the neighboring contact lines. This provides an important simplification: the liquid contact angles at the outer contact lines of each drop are essentially unaffected by the presence of the second drop, and the slow relaxation of the viscoelastic solid ensures circular liquid - vapor interface. As a result, the liquid contact angles at the interacting contact lines also remain constant to satisfy the circular cap shape of the liquid-vapor interface. In the example of fig.~\ref{expfig}(b) this liquid contact angle with respect to horizontal ($\theta_{\ell}$) is $\pi/2$ by symmetry, since we assume $\gamma_{sv}=\gamma_{s\ell}=\gamma_s$.

\begin{figure}[t]
\includegraphics[width=85mm]{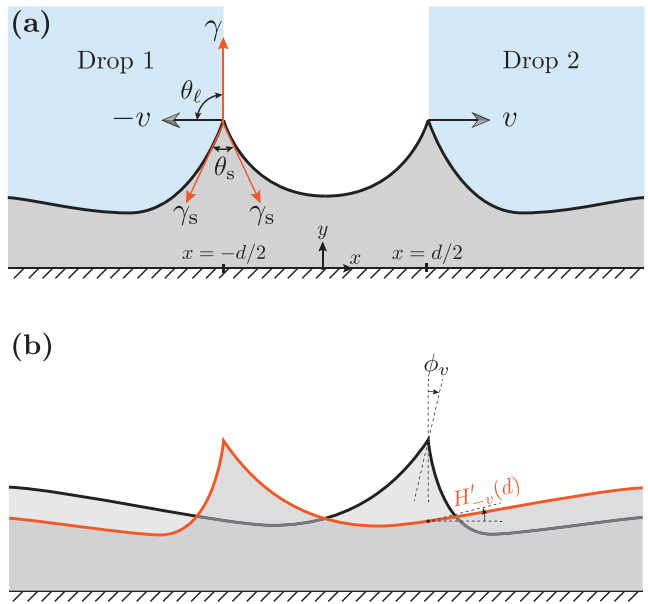}
\caption{Illustration of the interaction mechanism. (a) Schematic representation of the deformed substrate due to a repulsive interaction between two contact lines. (b) Dynamic deformation profiles associated to each of the contact line. Superposition of these two profiles give the resultant deformation.}
\label{int_fig}
\end{figure}

To understand the interaction mechanism, we magnify the deformation around neighbouring contact lines in fig.~\ref{int_fig}(a). It shows a sketch of two repelling contact lines with velocity $v$. Our goal is to find the value of $v$ for a given distance $d$. Deformation of the substrate at the three phase contact line is fully defined by the balance of three respective surface tensions, known as the Neumann condition \cite{MDSA12b,LUUKJFM14,Park2014}. The assumption of $\gamma_{sv}=\gamma_{s\ell}=\gamma_s$ leads to the symmetric wetting ridges of fig.~\ref{int_fig}(a). As demonstrated in the next section, in first approximation the solid profile is obtained by the superposition of the deformations due to the two moving contact lines shown in fig.~\ref{int_fig}b. However, motion on a viscoelastic medium causes a rotation of the wetting ridge by an angle $\phi_v$ that depends on the velocity: the larger the velocity, the larger the rotation angle \cite{Karpitschka2015}. At the same time, superposition results in a rotation of a wetting ridge due to the local slope of the other deformation profile ($H'_{-v}(d)$) at that location. Hence, the force balance with the capillary traction from an unrotated liquid interface can be maintained when both rotations exactly cancel each other. Therefore, by equating $H'_{-v}(d)$ to $\phi_v$, we obtain a relation between velocity and distance.
 
\subsection{Formulation}
In this section, we calculate the dynamic deformation of a linear, viscoelastic substrate below a moving contact line. This step will enable us to quantify the interaction velocity, according to the procedure described above. 

Due to the time-dependent relaxation behaviour of viscoelastic materials, the response of the substrate requires a finite time to adapt to any change in the imposed traction at the free surface. Under the assumption of linear response, the stress-strain relation of a viscoelastic solid is given by \cite{ferry1961}

\begin{equation}
\bm{\sigma}(\bm{x},t)=\int_{-\infty}^t\id t'\Psi(t-t')\dot{\bm{\epsilon}}(\bm{x},t')-p(\bm{x},t)\bm{I}.
\label{stsn1}
\end{equation}Here $\bm{\sigma}$ is the stress tensor, $\bm{\epsilon}$ is the strain tensor, $\Psi$ is the relaxation function, $p$ is the isotropic part of the stress tensor, $\bm{I}$ is the identity tensor, and $\bm{x}=\{x,y\}$. Throughout we will assume plain strain conditions, for which all strain components normal to the page vanish. We are interested in the surface profile of the deformed substrate, given by the vertical ($y$)-component of displacement vector $\bm{u}$. For small deformations this imply $u_y(x,h_0,t)=h(x,t)$. The soft substrate is attached to a rigid support at its bottom surface and subjected to traction $T(x,t)$ on the top surface due to the pulling of the contact line. In addition, the surface tension of the solid $\gamma_s$ provides a solid Laplace pressure that acts as an additional traction on the substrate. For inertia-free dynamics, the Cauchy equation along with the boundary conditions for the plane strain problem are given by

\begin{subequations}\label{odenbc}
\begin{align}
&\bm{\nabla}.\hspace{1mm}\bm{\sigma}=0,\\
&\left.\sigma_{yy}\right |_{y=h_0}=T(x,t)+\gamma_s\pdd{h}{x},\\
&\left.\sigma_{yx}\right |_{y=h_0}=0,\\
&\hspace{1mm}\bm{u}(x,0,t)=0.
\end{align}
\end{subequations} 
Following previous works \cite{LAL96,Karpitschka2015}, we solve eqns.~\eqref{stsn1} and~\eqref{odenbc} using a Green's function approach. Applying Fourier transforms in both space and time ($x\rightarrow q$ noted by `\~{}', $t\rightarrow\omega$ noted by `\^{}'), the deformation can be written as,

\begin{equation}
\widehat{\widetilde{h}}(q,\omega)=\widehat{\widetilde{T}}(q,\omega)\left[\gamma_s q^2+{\mu(\omega)\over\widetilde{\mathcal{K}}(q)}\right]^{-1}.
\label{hqw}
\end{equation}Forward and backward Fourier transforms (with respect to space) of any function is defined as, $\widetilde{f}(q)=\int_{-\infty}^{\infty}f(x)\e^{-\ii q x}\id x$ and $f(x)=\int_{-\infty}^{\infty}\widetilde{f}(q)\e^{\ii q x}\id q/2\pi$. Transformation rules are equivalent for $t\leftrightarrow\omega$. Here $\mu(\omega)$ is the complex shear modulus of the material and consists of the storage modulus $G'(\omega)=\Re[\mu(\omega)]$, and loss modulus $G''(\omega)=\Im[\mu(\omega)]$:

\begin{equation}
\mu(\omega)=\ii\omega\int_0^{\infty}\Psi(t)\e^{-\ii\omega t}\id t=G'(\omega)+\ii G''(\omega).
\label{mu}
 \end{equation}
The spatial dependence is governed by the kernel $\widetilde{\mathcal{K}}(q)$, which for an incompressible, elastic layer of thickness $h_0$ reads

\begin{equation}
\widetilde{\mathcal{K}}(q)=\left[\frac{\sinh(2qh_0)-2qh_0}{\cosh(2qh_0)+2(qh_0)^2+1}\right]{1\over 2q}.
\label{kq}
\end{equation}
Two backward transforms of~\eqref{hqw} give the surface deformation $h(x,t)$.

In what follows, we perform the calculations assuming a power-law rheology of the soft substrate, characterised by a complex modulus,
\begin{equation}
\mu(\omega)=G\left[1+(\ii\omega\tau)^n\right].
\label{muw}
\end{equation}
This rheology provides an excellent description of the reticulated polymers like PDMS, as are used in experiments on soft wetting \cite{JXWD11,durotaxisPNAS13,Karpitschka2015,Style13}. In particular, the loss modulus has a simple power-law form, ${G}'' \sim G(\omega \tau)^n$, where $G$ is the static shear modulus and $\tau$ is the relaxation timescale. The exponent $n$ depends on the stoichiometry of the gel, and varies between 0 and 1 \cite{winter86,scanlan91,chambon1987linear}. The experimental data presented in fig.~\ref{expfig} is for a PDMS gel with a measured exponent of $n=0.61$ and timescale $\tau=0.68s$. 
\subsection{Two steadily moving contact lines}

Now we turn to two moving contact lines, as depicted in fig.~\ref{int_fig}(a). Laplace pressure of the liquid drops scales with $R^{-1}$, and vanish in this limit of large drops. Hence, the applied traction is described by two delta functions (noted $\delta$), separated by a distance $d(t)$:

\begin{equation}
T(x,t)=\gamma\delta\left(x+{d(t)\over 2}\right)+\gamma\delta\left(x-{d(t)\over 2}\right).
\label{trac}
\end{equation}
Owing to the linearity of the governing equations, the deformations form a simple superposition

\begin{equation}
h(x,t)=h_1(x,t)+h_2(x,t).
\label{hxt}
\end{equation} 
Here $h_1(x,t)$ and $h_2(x,t)$ respectively are the solutions obtained for contact lines moving to the left and right.

Before considering the superposition, let us first discuss the deformation due to a contact line centered at the origin and moving steadily with velocity $-v$. The traction and deformation are then of the form $\gamma\delta(x+vt)$, and $H_{-v}(x+vt)$. When computing the shape in a co-moving frame, it can be evaluated explicitly from (\ref{hqw}) as 

\begin{equation}
\widetilde{H}_{-v}(q)=\gamma\left[\gamma_s q^2+{\mu(vq)\over\widetilde{\mathcal{K}}(q)}\right]^{-1}.
\label{h1q}
\end{equation}
The effect of velocity is encoded in the argument of the complex modulus $\mu$, and hence couples to the dissipation inside the viscoelastic solid. Setting $v=0$, this expression (\ref{h1q}) gives a perfectly left-right symmetric deformation of the contact line. As argued, for $v\neq 0$, the motion breaks the left-right symmetry and induces a rotation of the wetting ridge. 

Now, for two contact lines located at $\pm d(t)/2$ moving quasi-steadily with opposite velocities $v=\pm \frac{1}{2} \dot{d}$, the deformation in eq.~\eqref{hxt} can be written as

\begin{equation}
h(x,t)=H_{-v}\left(x+{d\over 2}\right)+H_{v}\left(x-{d\over 2}\right).
\label{hxt1}
\end{equation}
Under this assumption of quasi-steady motion, there is no explicit time-dependence; the shape is fully known once the velocity $v$ and the distance $d$ are specified. However, as discussed in sec.~\ref{sec:formulation}A, the total deformation of eq.~\eqref{hxt1} should be consistent with a liquid angle of $\pi/2$. This is satisfied when the ridge rotation due to motion, $\phi_v$, perfectly balances  the slope induced by the second drop. Summarizing this condition as

\begin{equation}
H'_{-v}(d)=\phi_v,
\label{inteqn}
\end{equation}
we can compute the velocity $v$ for given separation $d$ of the two contact lines. Explicit expressions can now be found from equation~\eqref{h1q}, as

\begin{equation}
H'_{-v}(d)=\int_{-\infty}^{\infty}\ii q\widetilde{H}_{-v}(q)\e^{\ii q d}{\id q\over 2\pi},
\label{h1x}
\end{equation}
and (See `Methods' section of \cite{Karpitschka2015} for details)
\begin{equation}
\begin{split}
\phi_v&=-{1\over 2}\left[H'_v\left(0^+\right)+H'_v\left(0^-\right)\right]\\
&=-\int_{-\infty}^{\infty}\Re\left[\ii q\widetilde{H}_v(q) \right]{\id q\over 2\pi}.
\label{phi}
\end{split}
\end{equation} 
In what follows we numerically solve eq.~\eqref{inteqn}, and obtain the interaction velocity as a function of $d$. 
 
\subsection{Dimensionless form}

The results below will be presented in dimensionless form. We use the substrate thickness $h_0$ as the characteristic lengthscale, relaxation timescale of the substrate $\tau$ as the characteristic timescale, and introduce the following dimensionless variables:
\begin{equation}
\begin{split}
&\overbar{q}=qh_0,\hspace{2mm}\overbar{x}={x\over h_0},\hspace{2mm}\overbar{H}={H\over h_0},\hspace{2mm}\bar{d}={d\over h_0},\hspace{2mm}\bar{v}={v\tau\over h_0},\\
&\hspace{2mm}\overbar{\omega}=\omega\tau,\hspace{2mm}\overbar{\widetilde{\mathcal{K}}}={\widetilde{\mathcal{K}}\over h_0},\hspace{2mm}\overbar{\widetilde{H}}={\widetilde{H}\over h_0^2},\hspace{2mm}\overbar{\mu}={\mu\over G}.
\label{ndvars}
\end{split}
\end{equation}
In the remainder of the paper we drop the overbars, unless specified otherwise. Eqns~\eqref{inteqn},~\eqref{h1x},~\eqref{phi} remain unaltered in dimensionless form, but the scaled deformation reads
\begin{equation}
\widetilde{H}_{-v}(q)=\alpha\left[\alpha_s q^2+{\mu(vq)\over\widetilde{\mathcal{K}}(q)}\right]^{-1},
\label{Hqv}
\end{equation}
This form suggests that the two main dimensionless parameters of the problem are, 

\begin{equation}
\alpha={\gamma\over Gh_0}, \quad \quad \alpha_s={\gamma_s\over Gh_0},
\end{equation}
which can be seen as two independent elastocapillary numbers. $\alpha$ compares the typical substrate deformation to its thickness, and as discussed in the next section $\alpha_s$ measures the decay of substrate deformation compared to $h_0$. The assumption of linear (visco)elasticity demands that the typical slope of the deformation is small, which requires $\alpha/\alpha_s = \gamma/\gamma_s \ll1$. In the final section we will also consider drops of finite size $R$. In that case, the ratio $R/h_0$ appears as a third dimensionless parameter.

\section{Results}\label{sec:results}

\subsection{Relation between velocity and distance}

\begin{figure}
\includegraphics[width=85mm]{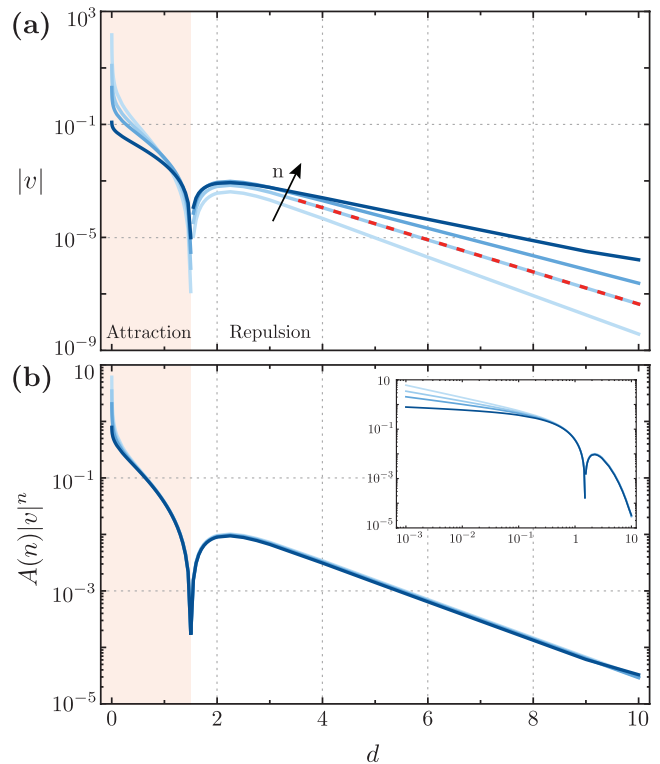}
\caption{(a) velocity-distance plots for increasing n values (n=0.5, 0.6, 0.7, 0.9), for the parameter $\alpha_s=1$. The red dashed line represent the far-filed asymtotic given in eq.~\eqref{vvsdasymp} for $n=0.7$. (b) Rescaling the velocity gives a perfect collapse for small velocity (large distance). Inset: rescaled velocity vs distance on a log-log plot. The prefactor $A(n)$ is given by $A(n)=n2^{n-1}/\cos(n\pi/2)$.}
\label{vvsd}
\end{figure}

Here we discuss our main result, namely the contact line velocity $v$ of the inverted Cheerios effect, as a function of distance $d$. Figure~\ref{expfig}(d) represents a typical velocity-distance plot for $\alpha_s=1$ and $n=0.7$ obtained by numerically solving eq.~\eqref{inteqn}. The theory shows qualitative agreement with the experiments: the interaction velocity is repulsive at large distance, but displays an attraction at small distances. The change in the nature of interaction can be understood from the local slope of deformation. As can be seen from fig.~\ref{int_fig}(b), the slope $H'_{-v}$ is negative at small distances from the contact line. This induces a negative velocity (attraction) to the contact line located at $d/2$. However, incompressibility of the material causes a dimple around the wetting ridge, and at $d\sim 1$ the slope changes its sign and so does the interaction velocity. A fully quantitative comparison between theory and experiment, taking into account finite drop size effects, will be presented at the end of the paper in Sec.~\ref{discuss}. 

We now study the velocity-distance relation in more detail. In fig.~\ref{vvsd}(a), we show the interaction velocity on a semi-logarithmic scale, for different values of the rheological parameter $n$. In all cases the interaction is attractive at small distance, and repulsive at large distance. The point where the interaction changes sign is identical for all these cases. The straight line at large $d$ reveals that the interaction speed decays exponentially, $v\sim\e^{-\lambda d}$. The exponent $\lambda$ is not universal, but varies as $\lambda\sim1/n$. Indeed by rescaling the velocity as $|v|^n$, we observed a collapse of the curves (cf. fig.~\ref{vvsd}(b)) in the ranges where the interaction velocities are small. The inset of fig.~\ref{vvsd}(b), however, shows that the collapse does not hold at very small distances where the velocities are larger.  

To understand these features, we extract the far-field asymptotics of the velocity-distance curve. As $v$ decreases for large $d$, we first expand eq.~\eqref{h1x} for small $v$ to get

\begin{equation}
H'_{-v}\simeq\alpha\int_{-\infty}^{\infty}\ii q\left[\alpha_s q^2+{1\over\widetilde{\mathcal{K}}(q)}\right]^{-1}\e^{\ii q d}{\id q\over 2\pi}+\mathcal{O}(v).
\label{hxasymp}
\end{equation}
This effectively corresponds to the slope of a stationary contact line and the viscoelastic effects drop out. Hence, at large distances away from the contact line the deformed shape is essentially static. The substrate elastocapillary number ($\alpha_s$) governs the deformed shape in far field. In the Appendix we evaluate this static slope, which yields 
\begin{equation}
H'_{-v}\simeq\alpha\hspace{1mm}C(\alpha_s)\hspace{1mm}\e^{-\widetilde{\lambda}(\alpha_s) d},
\label{hxsmallv}
\end{equation}
indeed providing an exponential decay. This relation confirms that $\alpha_s$ governs the decay of substrate deformation. Similarly, we expand $\phi_v$ in eq.~\eqref{phi} for small $v$ as (see \cite{Karpitschka2015}), 
\begin{equation}
\phi_v\simeq\alpha\frac{n 2^{n-1}}{\alpha_s^{n+1}\cos{n\pi\over2}}v^n.
\label{phismallv}
\end{equation} 
Hence, the rotation of the ridge inherits the exponent $n$ from the rheology. Comparison of eqs~\eqref{hxsmallv} and~\eqref{phismallv} gives the desired asymptotic relation 
\begin{equation}
v=\left[{\alpha_s^{n+1}C(\alpha_s)\over A(n)}\right]^{1\over n}\e^{-{\widetilde{\lambda}(\alpha_s)\over n}d},
\label{vvsdasymp}
\end{equation}
with $A(n)=n2^{n-1}/\cos(n\pi/2)$. For $\alpha_s=1$, $\widetilde{\lambda}=0.783$ and $C=0.069$. The red dashed line in fig.~\ref{vvsd}(a) confirms this asymptotic relation. The quantity $A(n)v^n$ is independent of the substrate rheology and collapses to a universal curve for large distances. Only at small separation, a weak dependency on $n$ appears (see the inset of fig.~\ref{vvsd}(b)).


 \begin{figure} 
\includegraphics[width=85mm]{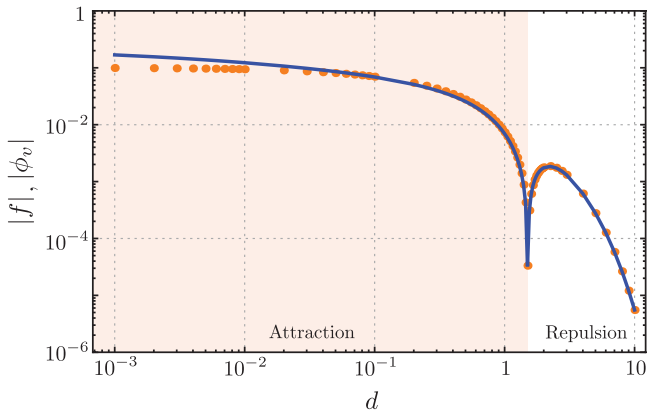}
\caption{Force vs distance plot for $\alpha=1/5$ and $\alpha_s=1$. The blue line represents the dissipative drag force for $n=0.7$ given in eq.~\eqref{phi}. The red points represent the localized force given by eq.~\eqref{fint}. Beyond small $d$, the slope of the dynamic profile is similar its static counterpart. As a result, two forces agree well.}
\label{fvsd}
\end{figure}

 \begin{figure*}
\includegraphics[width=175mm]{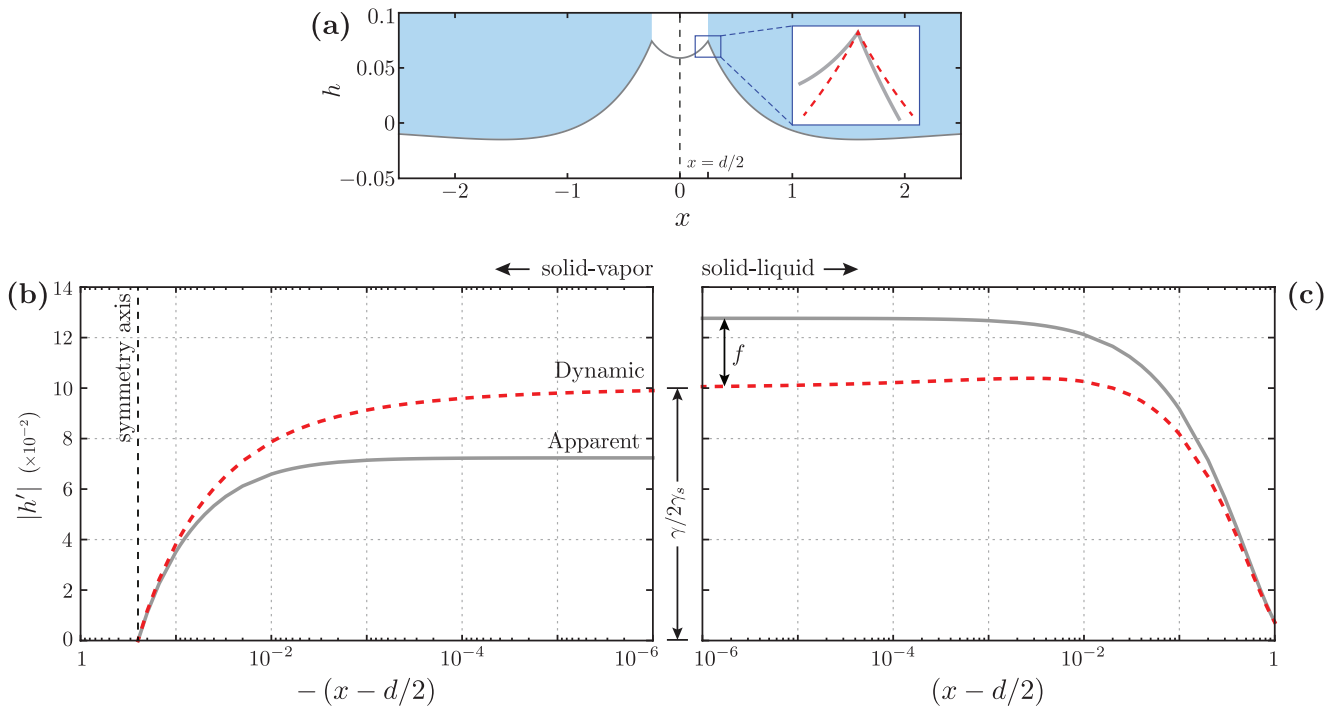}
\caption{(a) Resultant deformation in the localized-force model, obtained by superimposing two static profiles ($d=1/2$, for $\alpha=1/5$ and $\alpha_s=1$). The gray curve represents a resultant static profile. The red dashed curve shows the corresponding dynamic wetting ridge that maintains equilibrium shape. (b), (c) Slope of the resultant deformation profiles (dynamic and static) measured away from the contact line at $x=d/2=1/4$ in both directions. The red dashed lines are the dynamic slopes, approaching $\gamma/2\gamma_s$ at the contact line. The gray lines correspond to the localized-force model, showing an apparent contact angle that differs by an amount $f$.}
\label{slopevsx}
\end{figure*}

\subsection{Force-distance}

We now interpret these velocity-distance results in terms of an interaction force, mediated by the substrate deformation. The underlying physical picture is that the interaction force between the two contact lines induces a motion to the drops. In a regime of overdamped dynamics, dissipation within the liquid drop can be neglected, and the interaction force must be balanced by a dissipative force due to the total viscoelastic drag inside the substrate. In fact, eq.~\eqref{inteqn} can be interpreted as a force balance: the slope $H'_{-v}(d)$ induced by the other drop is the interaction force, while the dissipative force is given by the rotation $\phi_v$ due to drop motion. Viscoelastic dissipation given by the relation $G''(\omega)\sim\omega^n$ manifests itself in the ridge rotation through, $\phi_v\sim v^n$. The blue line in fig.~\ref{fvsd}, shows the corresponding dissipative force as a function of distance for $n=0.7$. 

In Ref.~\cite{Karpitschka2016} we presented a quasi-static approach to compute the interaction force. There, the main hypothesis was that the dissipative force can be modelled as a point force $f$ that is perfectly localized at the contact line. We are now in the position to test this hypothesis by comparing its predictions to those of the dynamical theory developed above, where the viscoelastic dissipation inside the solid is treated properly. In the localized-dissipation model, the deformation profiles are treated as perfectly static and thus involve $H'_{v=0}(d)$. However, in this approximation the change in deformation profile now leads to a local violation of the Neumann balance (cf. inset figure~\ref{slopevsx}(a)), since the static model has no ridge rotation, $\phi_{v=0}=0$. This imbalance in the surface tensions must be compensated by the localized dissipative force $f$, which can subsequently be equated to the interaction force. Hence, for the case of large drops, the localized-force model predicts

\begin{equation}
\overbar{f}=f/\gamma=H'_{v=0}(d).
\label{fint}
\end{equation}
Like before, we drop the overbar and denote the dimensionless force by $f$. The result of the localized-force model is shown as the red data points in fig.~\ref{fvsd}, showing $f$ versus $d$. At large distance this gives a perfect agreement with the dynamical theory. For large $d$, $H'_{-v}(d)\simeq H'_{v=0}(d)$, and the dissipative force $\phi_v$ perfectly agrees with the interaction force $f$. In this limit, slope of the deformation decays exponentially with distance, and so is the interaction force. However, as $d\rightarrow0$, the static profiles differ from the dynamic profiles, inducing a small error in the calculation of the interaction force.  

The differences between the localized-dissipation model and the full viscoelastic dynamic calculation can be interpreted in terms of true versus apparent contact angles. Figure~\ref{slopevsx}(a) shows the profile in the approximation of the localized-force model. The inset shows the violation of the Neumann balance: the red dashed line is the dynamic profile obtained from our viscoelastic theory. A more detailed view on the contact angles is given in fig.~\ref{slopevsx}(b,c). Here we compare the dynamical slope (red dashed line) to that of the localized-force model (gray solid line), on both sides of the contact line at $x=d/2$. The two profiles have the same slope beyond $x\sim\gamma_s/Gh_0$. However, differences appear at small distance near the contact line where viscoelastic dissipation plays a role. While the dynamic profiles are left-right symmetric, the localized-force model gives a symmetry breaking: the difference in slope between the two approach at the contact line gives the magnitude of $f$. 

In summary, the inverted Cheerios effect can be characterised by an interaction force, which at separations ($d$) larger than $\gamma_s/Gh_0$ is independent of substrate rheology. In this regime the interaction force can be computed assuming that all dissipation is localized at the contact line -- this gives rise to an apparent violation of the Neumann balance, due to a dissipative force $f$. In reality the Neumann balance is restored at small distance, but the total dissipative force is indeed equal to $f$. At separations $d < \gamma_s/Gh_0$, however, the spatial distribution of dissipation has significant effects and the localized-force model gives rise to some (minor) quantitative errors. 

\section{Drops of finite size}\label{sec:radius}

We now show that interaction is dramatically changed when considering drops of finite size $R$, compared to the substrate thickness $h_0$. A fully consistent dynamical theory is challenging for finite-sized drops, since the two contact lines of a single drop are not expected to move at the same velocity. However, the previous section has shown that the localized-force model provides an excellent description of the interaction force, which becomes exact at separations $d>\gamma_s/Gh_0$.  We therefore follow this approximation to quantify the inverted Cheerios effect for finite sized drops. 

\subsection{Effect of substrate thickness}

\begin{figure} 
\includegraphics[width=85mm]{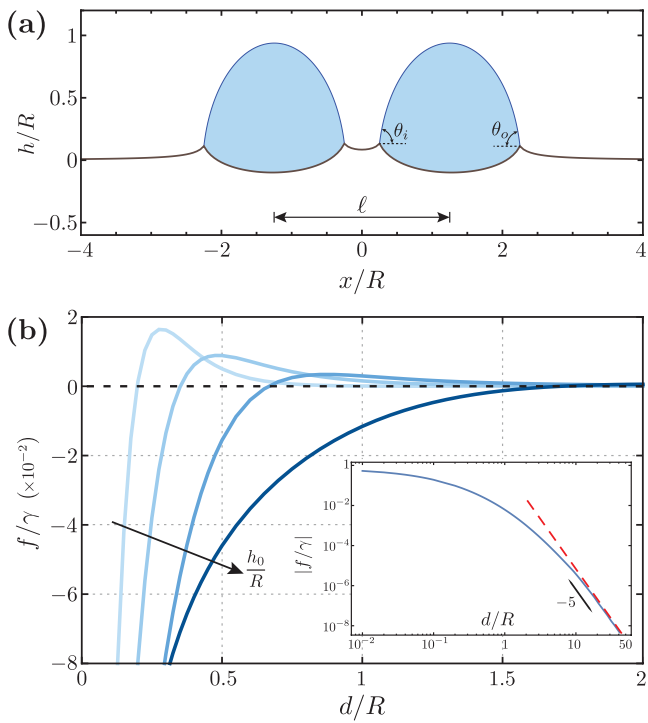}
\caption{(a) Substrate deformation and liquid cap shape due to two interacting drops centered at $x=\pm5/4$. The separation distance between two neighboring contact lines is $d/R=\ell/R-2=1/2$. (b) The interaction force vs distance plot as the substrate thickness to drop radius ratio is increased $(h_0/R=1/10, 1/5, 2/5, 1)$. For thick substrates the interaction becomes predominantly attractive. Inset: Force vs distance plot on log scale for finite drops of an elastic half space. The red dashed line is the far-field asymptotic representing $f\sim d^{-5}$. The parameters used in this figure are $\gamma/GR=2/5$, and $\gamma_s/GR=1/5$. $\theta_{eq}$ for this case is found to be $78.91^{\circ}$.}
\label{dfd}
\end{figure}

We assume that during interaction, radius of both the drops does not change and the liquid cap remains circular (i.e. no significant dissipation occurs inside the liquid). The traction applied by a single drop on the substrate comprised of two localized forces pulling up at the contact lines and a uniform Laplace pressure pushing down inside the drop. The traction applied by a single drop of size $R$, centered at the origin, is thus given by

\begin{equation}
\begin{split}
&T_d(x)=\gamma\sin\theta_{eq}\left[\delta(x+R)+\delta(x-R)\right]\\
&-\gamma{\sin\theta_{eq}\over R}\Theta(R-|x|),
\label{}
\end{split}
\end{equation}
expressed in dimensional form. Here $\Theta$ is the Heaviside step function. The corresponding Fourier transform reads 
\begin{equation}
\widetilde{T}_d(q)=2\gamma\sin\theta_{eq}\left[\cos (qR)-{\sin(qR)\over qR}\right].
\label{tdq}
\end{equation}
Considering two drops centered at $x=-\ell/2$ and $x=\ell/2$ respectively, with corresponding tractions of $T_d(x+\ell/2)$ and $T_d(x-\ell/2)$, the total traction induced by the two drops reads, in Fourier space

\begin{equation}
\widetilde{T}(q)=\widetilde{T}_d(q)\e^{\ii q\ell/2}+\widetilde{T}_d(q)\e^{-\ii q\ell/2},
\label{ttd}
\end{equation}
where the inter-drop separation $d=\ell-2R$. Using the static Green's function approach, we compute the corresponding deformation profiles. A typical profile obtained from this traction is shown in fig.~\ref{dfd}. 

The problem is closed by determining the liquid contact angles at both the contact lines ($\theta_i$, $\theta_o$) self-consistently with the elastic deformation \cite{LUUKJFM14}. The interaction is found from reestablishing a violated Neumann balance by a dissipative force (for details we refer to \cite{Karpitschka2016}). As such we determine the force-distance relation for varying drop sizes. The results are shown in fig.~\ref{dfd}(b), for various drop sizes $h_0/R$. Clearly, upon decreasing the drop size, or equivalently increasing the layer thickness, the interaction force becomes predominantly attractive. For a thin substrate, incompressibility leads to a sign change of the interaction force. As the substrate thickness is increased the profile minimum is moved away from the contact line. For an elastic half space, i.e. $h_0/R=\infty$, the deformation monotonically decays away from the contact line giving rise to a purely attractive interaction. The inset of fig.~\ref{dfd}(b) shows the interaction force for an elastic half space.

\subsection{Far-field asymptotics on a thick substrate}

When two drops are far away from each other, $\theta_i=\theta_o\simeq\theta_{eq}$ and the liquid cap resembles its equilibrium shape. In this limit, the drop-drop interaction is mediated purely by the elastic deformation. Here we calculate the large-$d$ asymptotics of the interaction force. We follow a procedure similar to that proposed for elasticity-mediated interactions between cells \cite{SchwartzPRL2002,BischofsPRE2004}, where the cells are treated as external tractions. The total energy of the system $E_{tot}$ then consists of the elastic strain energy $E_{e\ell}$ plus a term $E_w$ originating from the work done by the external traction. At fixed separation distance, the mechanical equilibrium of the free surface turns out to give $E_{e\ell}=-\frac{1}{2}E_w$ (the factor 1/2 being a consequence of the energy in linear elasticity). So, we can write $E_{tot} = E_{e\ell} + E_w = - E_{e\ell}$ \cite{BischofsPRE2004}. Subsequently, the interaction force between the droplets is computed as $f=-\d{E_{tot}}{d}= \d{E_{e\ell}}{d}$. The task will thus be to evaluate $E_{e\ell}$ from the traction induced by two drops.

In the absence of body forces, the elastic energy (per unit length) is written as

\begin{equation}
\begin{split}
E_{e\ell}&={1\over2}\int_{-\infty}^{\infty}T(x)h(x)\id x\\
&={1\over2G}\int_{-\infty}^{\infty}\widetilde{T}(q)\widetilde{T}(-q)\widetilde{K}(q){\id q\over2\pi}.
\label{Eel}
\end{split}
\end{equation}
Plugging in the traction applied by two drops given in eq.~\eqref{ttd} into eq.~\eqref{Eel} we find that there are two parts of the elastic energy: one is due to the work done by individual tractions (the self-energy of individual drops), and an interaction term ($E_{int}$) given by  

\begin{equation}
E_{int}={1\over G}\int_{-\infty}^{\infty}\widetilde{T}_d(q)\widetilde{T}_d(-q)\widetilde{K}(q)\left(\e^{\ii q\ell}+\e^{-\ii q\ell}\right){\id q\over4\pi}.
\label{}
\end{equation}

To obtain the far-field interaction energy ($\ell \gg R$), we now perform a multipole expansion of the traction $T_d(x)$, or equivalently a long-wave expansion of its Fourier transform. The quadrupolar traction of eq.~\eqref{tdq} gives, after expansion

\begin{equation}
\widetilde{T}_d(q)\simeq{(-\ii q)^2\over2}Q_2,
\label{}
\end{equation}
were $Q_2$ is the second moment of the traction and is given by, $Q_2=\int_{-\infty}^{\infty}T_d(x)x^2\id x={4\over3}\gamma\sin\theta_{eq}R^2$. So, the interaction energy in the far-field reads

\begin{equation}
\begin{split} 
E_{int}&\simeq{1\over G}\int_{-\infty}^{\infty}{(\ii q)^4\over2^2}Q_2^2\widetilde{K}(q)\left(\e^{\ii q\ell}+\e^{-\ii q\ell}\right){\id q\over4\pi}\\
&={Q_2^2\over8G}\left(K''''(\ell)+K''''(-\ell)\right).
\label{}
\end{split}
\end{equation}
For drops on a finite thickness, the Green's function $K(x)$ decays exponentially and again gives an exponential cut-off of the interaction at distances beyond $h_0$. 

A more interesting result is obtained for an infinitely thick substrate, formally corresponding to $h_0 \gg \ell$. The kernel for a half space is $K(x)=-{\log|x|\over2\pi}$, which can e.g. be inferred from the large thickness limit of (\ref{kq}). Consequently, $K''''(\ell)=K''''(\ell)={3\over\pi\ell^4}$, so that the interaction energy becomes 

\begin{equation}
E_{int}={4\over3\pi} \frac{(\gamma \sin \theta_{eq})^2}{G}\frac{R^4}{\ell^4}.
\label{eq:bloeb}
\end{equation} 
This is identical to the interaction energy between capillary quadrupoles \cite{Stamou2000,Danov2005,Danov2010}, except for the prefactor that depends on elastocapillary length in the present case. We calculate the force of interaction by $f=\d{E_{int}}{\ell}$, giving

\begin{eqnarray}
\frac{f}{\gamma}&=&-{16 \sin^2 \theta_{eq}\over3\pi} \left(\frac{\gamma }{GR}\right) \left(\frac{R}{\ell}\right)^5 \nonumber \\
&\simeq& -{16 \sin^2 \theta_{eq}\over3\pi} \left(\frac{\gamma }{GR}\right) \left(\frac{R}{d}\right)^5,
\label{eq:bloeb2}
\end{eqnarray} 
where the result is collected in dimensionless groups. 
Here we replaced $\ell \simeq d$, which is valid when the drop separation is large compared to $R$. The red dashed line in the inset of fig.~\ref{dfd}(b) confirms the validity of the above asymptotic result, including the prefactor. Hence, in the limit of large thickness the attractive interaction force decays algebraically, as $1/d^5$, which is in stark contrast to the exponential decay observed for thin substrates. 

It is instructive to compare  eq.~\eqref{eq:bloeb} with the interaction energy for capillary quadrupoles formed by rigid anisotropic particles embedded in a fluid-fluid interface \cite{Stamou2000,botto2012}. In that case an undulated triple line is formed owing to contact line pinning \cite{Stamou2000} or to the change in curvature of the solid surface along the triple line \cite{loudet2006, Lewan10}. For capillary quadrupoles the far-field interaction energy is proportional to $ \gamma H_p^2 (R/\ell)^4$ , where $H_p$ is the amplitude of the quadrupolar contact line distortion.   Despite an identical power-law dependence, the prefactors in eq. \ref{eq:bloeb} and in the corresponding expression for capillary quadrupoles are different, reflecting a different physical origin.  A capillary quadrupole is formed by four lobes located at an average distance R from the quadrupole center: two positive meniscus displacements of amplitude $H_p$ along one symmetry axis of the particle, and two negative displacements along the perpendicular axis. The presence of a second particle changes the slope of equal sign meniscii along the center-to-center line, resulting in a net capillary attraction. The corresponding interaction force per unit length of contact line is proportional to $\gamma (H_p/R)^2 (R/d)^5$, to be compared with eq. \ref{eq:bloeb2}.  The non-dimensional ratio $(H_p/R)$ is a geometric feature of the particle that depends primarily on the particle shape and contact angle \cite{botto2012}. This ratio is  independent of the particle size. On the contrary, the non-dimensional ratio $\frac{\gamma }{GR}$ does depend on the particle size, as well as on the material-dependent scale $\gamma/G$. A differentiating feature is also that in the elastocapillary quadrupole the upward and downward distortions of the elastic surface are located along the line of interaction,  thus forming a linear quadrupole. The identical power-law dependence for capillary and elasto-capillary quadrupoles  is incidental. It is a consequence of the fact that the $\log$ function is the Green's function for both the small-slope Young-Laplace equation of capillarity and the equation governing 2D linear elasticity.

\begin{figure*}
\includegraphics[width=175mm]{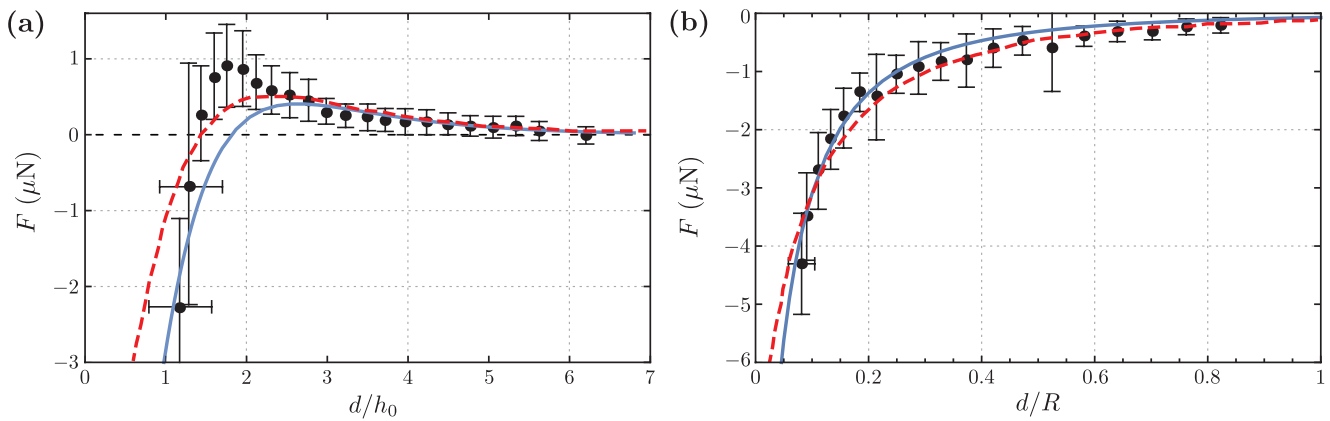}
\caption{Comparison between theory and experiment. The experimental data (black points) and 3D theory (red dashed lines) is reproduced from \cite{Karpitschka2016}. The drops have an average radius ($R$) of 700 $\mu$m in both cases. The liquid (ethylene glycol) and the solid (PDMS) surface tensions are $\gamma=48$ mN/m, and $\gamma_s\simeq20$ mN/m respectively. Shear modulus of the substrate is $G=280$ Pa. The blue lines are the interaction force obtained from 2D theory. (a) Attraction and repulsion on a substrate of thickness ($h_0$) $40$ $\mu$m. (b) Purely attractive interaction between drops on a thick substrate ($h_0=8$ mm). The adjustable length, multiplied with the 2D interaction force for comparison with the experiments is $\pi R/6$ for both the cases.}
\label{theorynexp}
\end{figure*}

\section{Discussion}
\label{discuss}

Finally, let us turn to experiments of the inverted Cheerios effect. The results in fig.~\ref{expfig} already show a qualitative agreement between theory and experiment. A quantitative comparison in terms of ``velocity vs distance" is challenging, since the experimental motion follows is due to a combination of drop-drop interaction and gravity -- the latter was not taken into account in the dynamical theory. However, gravity is easily substracted when representing the result as ``force vs distance", as was also done in \cite{Karpitschka2016}. The experimental data are reproduced in fig~\ref{theorynexp}: On thick substrates the interaction is purely attractive (panel b), while a change in sign of the interaction force is observed on thin substrates (panel a). The  two-dimensional model presented in this paper predicts a force per unit length $f$. Hence, to compare to experiment one needs to multiply the result with a proper length over which this interaction is dominant. The blue lines are obtained by choosing a length of $\pi R/6$, i.e. a small fraction of the perimeter, giving a good agreement with experiment. In \cite{Karpitschka2016} we provided numerically a three-dimensional result, based on the same localized-force approach. The result is shown as the red lines in fig.~\ref{theorynexp} and is similar to the two-dimensional case. This good agreement with experiment shows that the elasto-capillary theory presented here provides an accurate description of the physics. The similarity between the two- and three-dimensional models shows that the interaction is dominated by the regions where the contact lines are closest.

In conclusion, we present a theoretical model to capture the dynamical interaction between two liquid drops on a soft, solid surface. For large drops the interaction is limited to neighboring contact lines, and we find the interaction velocity-distance relation in this limit. For large distances, the velocity decays exponentially with an exponent that depends both on substrate rheology and surface tension. The over damped dynamics of drop motion allow us to relate the velocity to an effective viscoelastic drag force at the contact line. We also connect our theory to a quasi static description of the interaction, quantified in terms of force-distance curves, and find that the two approaches agree well for large inter drop distances. Motivated by this, we extend the quasi-static force calculation to finite sized drops to show how the ratio of substrate thickness to drop radius governs the nature of the interaction. In the other physically relevant limit of small drops on a very thick substrate, the interaction force is found to decay algebraically for large distances. We anticipate that the dissipation mediated interaction mechanism studied here to play a key role in collective dynamics of drops on soft surfaces \cite{durotaxisPNAS13,Bueno2017}, and dynamics of soft, adhesive contacts. 

\appendix*
\section{}

We evaluate the complex counterpart of the integrand in eq.~\eqref{hxasymp} on a contour $C$, given by
\begin{equation}
\alpha\ointctrclockwise_C\ii \mathbf{q}\left[\alpha_s\mathbf{q}^2+{1\over\widetilde{K}(\mathbf{q})}\right]^{-1}\e^{\ii\mathbf{q}d}{\id \mathbf{q}\over2\pi}.
\label{complint}
\end{equation} Here $\mathbf{q}$ is the complex wave vector. The contour $C$ consists of a semicircle of radius $r$ centered at the origin and a straight line connecting the ends of the semicircle. We evaluate the above integral using Residue theorem. As $r\rightarrow\infty$, the integrand vanishes on the semicircle and the contour integral simplifies to eq.~\eqref{hxasymp}. For any $\alpha_s$ the integrand of eq.~\ref{complint} has an infinite number of poles on the complex plane. The leading order behavior of the integration is governed by the pole with smallest imaginary part ($\mathbf{q}_{pole}$) as other poles lead to much faster decay. 

\begin{figure}[h]
\vspace{5mm}
\includegraphics[width=85mm]{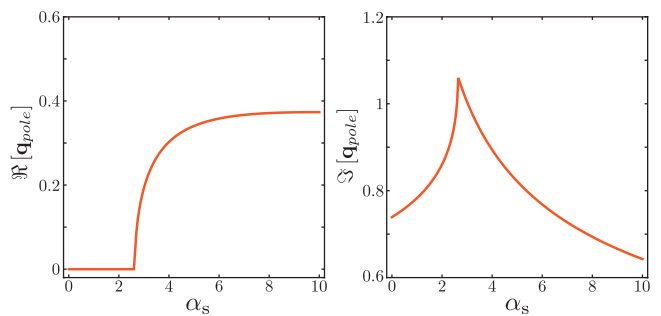}
\caption{Variation of $\mathbf{q}_{pole}$ on the complex plane with the elastocapillary number $\alpha_s$. Below a critical $\alpha_s$, $\mathbf{q}_{pole}$ is purely imaginary and signifies an exponential decay.}
\label{qpoles}
\end{figure}

The substrate elastocapillary number $\alpha_s$ dictates the value of $\mathbf{q}_{pole}$. Figure~\ref{qpoles}  shows that below a critical $\alpha_s$, $\mathbf{q}_{pole}$ is purely imaginary suggesting an purely exponential decay of the deformation. Beyond the critical $\alpha_s$, $\mathbf{q}_{pole}$ has a nonzero real and imaginary part giving rise to oscillatory behavior of slope as predicted by Long et. al \cite{LAL96}. Hence, the large distance approximation of slope is given by,
\begin{equation}
H'_{-v}\simeq 2\pi\ii \hspace{1mm}\mathrm{Res}\left[\frac{\alpha\ii\mathbf{q}}{\alpha_s \mathbf{q}^2+{1\over\widetilde{\mathcal{K}}(\mathbf{q})}}{\e^{\ii\mathbf{q}d}\over2\pi}\right]_{\mathbf{q}=\mathbf{q}_{pole}}.
\label{h'smallv}
\end{equation} This expression simplifies to eq.~\eqref{hxsmallv} in the main text.

\begin{acknowledgments}
L.B. acknowlwdges financial support from European Union Grant CIG 618335. S.K. acknowledges financial support from NWO through VIDI Grant No. 11304. A.P. and J.H.S. acknowledge financial support from ERC (the European Research Council) Consolidator Grant No. 616918.
\end{acknowledgments}


%

\end{document}